\documentclass{iau}
\usepackage{graphicx}

\title[Stable Magnetohydrodynamic Equilibria in Barotropic Stars?] 
{Search for Stable Magnetohydrodynamic Equilibria in Barotropic Stars.}

\author[J.P. Mitchell \& J. Braithwaite \& N. Langer \& A. Reisenegger \& H. Spruit]   
{J.P. Mitchell$^1$ $^2$ , J. Braithwaite$^2$ , N. Langer$^2$ , A. Reisenegger$^1$ \and H. Spruit$^3$}

\affiliation{$^1$Instituto de Astrof\'{i}sica, Facultad de F\'{i}sica, Pontificia Universidad Cat\'{o}lica de Chile, \\ Av. Vicu\~na Mackenna 4860 7820436 Macul, Santiago - Chile \\ email: {\tt jmitchel@astro.puc.cl} \\[\affilskip]
$^2$Argelander Institut, University of Bonn, \\ Auf dem Huegel 71, 53121 Bonn -  Germany \\[\affilskip]
$^3$Max-Planck-Institut f\"{u}r Astrophysik,\\ Karl-Schwarzschild-Str. 1, D-85748 Garching, Germany}

\pubyear{2013}
\volume{302}  
\pagerange{119--126}
\jname{Magnetic Fields Throughout Stellar Evolution}
\editors{A.C. Editor, B.D. Editor \& C.E. Editor, eds.}

\begin{document}

\maketitle

\begin{abstract}
It is now believed that magnetohydrodynamic equilibria can exist in stably stratified stars due to the seminal works of \cite[Braithwaite \& Spruit (2004)]{Braithwaite_2004} and \cite[Braithwaite \& Nordlund (2006)]{Braithwaite_2006}. What is still not known is whether magnetohydrodynamic equilibria can exist in a barotropic star, in which stable stratification is not present. It has been conjectured by \cite[Reisenegger (2009)]{Reisenegger_2009} that there will likely not exist any magnetohydrodynamical equilibria in barotropic stars. We aim to test this claim by presenting preliminary MHD simulations of barotropic stars using the three dimensional stagger code of \cite[Nordlund \& Galsgaard (1995)]{Nordlund_1995}.
\end{abstract}

\firstsection 
\section{Introduction}

The search for magnetohydrodynamic (MHD) equilibria in stars has been an ongoing study for the last few decades.  Early works looked at the stability of purely poloidal and purely toroidal magnetic fields in axial symmetry.  The configurations in each of these cases were found to be unstable.  The instability in the purely toroidal fields was found by \cite[Tayler (1973)]{Tayler_1973}, who showed that the field was prone to de-stabilization due to the kink and and interchange instabilities.  In the purely poloidal field case, it was found by \cite[Markey \& Tayler (1973)]{Markey_1973} and \cite[Wright (1973)]{Wright_1973} that the field is unstable in the region around the "neutral line", where the field vanishes.  It was then suggested that if MHD equilibria exist, they would consist of a mixed poloidal-toroidal field, as each component would work to stabilize the other.  Due to the difficulty of analytically solving the mixed poloidal-toroidal fields, an impasse was reached in the community until the numerical breakthrough of \cite[Braithwaite \& Spruit (2004)]{Braithwaite_2004} and \cite[Braithwaite \& Nordlund (2006)]{Braithwaite_2006}.  The authors used 3-D simulations of a stably stratified star to show that an initially random magnetic field could reach a twisted-torus MHD equilibrium, which was stable on timescales much longer than the Alfv\'{e}n time.  

The aforementioned works of \cite[Braithwaite \& Spruit (2004)]{Braithwaite_2004} and \cite[Braithwaite \& Nordlund (2006)]{Braithwaite_2006} had focused on stably stratified stars.  Whether or not stable MHD equilibria can exist in a barotropic star is still under debate.  It was suggested by \cite[Akgun et. al. (2013)]{Akgun_2013} that a key ingredient in the stability of the MHD equilibria in stably stratified stars is their stable stratification.  This stratification requires very strong forces in the radial direction in order to overcome the stratification.  Further conjectures by \cite[Reisenegger (2009)]{Reisenegger_2009} have claimed that stable MHD equilibria will not exist in barotropic stars.

On the other hand, there has been extensive work done looking for MHD equilibria in axially symmetric barotropic stars: \cite[Tomimura \& Eriguchi (2005)]{Tomimura_2005}, \cite[Ciolfi et. al. (2009)]{Ciolfi_2009}, \cite[Lander \& Jones (2009)]{Lander_2009}, \cite[Reisenegger (2009)]{Reisenegger_2009}, \cite[Akg{\"u}n et. al. (2013)]{Akgun_2013}, \cite[Armaza et. al. (2013)]{Armaza_2013}.  Although these works have found MHD equilibria, the stability of the equilibria have not been investigated and they have all been limited to the axially-symmetric case.  

To investigate the possible existence of stable MHD equilibria in barotropic stars, we have evolved initially random magnetic field configurations in barotropic stars to see if stable equilibria can be reached.  In Section 2 of the paper we discuss the models used in our simulations.  Section 3 contains our preliminary results in the search for stable MHD equilibria in barotropic stars, and Section 4 contains a discussion of our results and of future studies. 

\section{The Models} 
\begin{figure}[b]
\hfill
 (a){\includegraphics[width=2.2in]{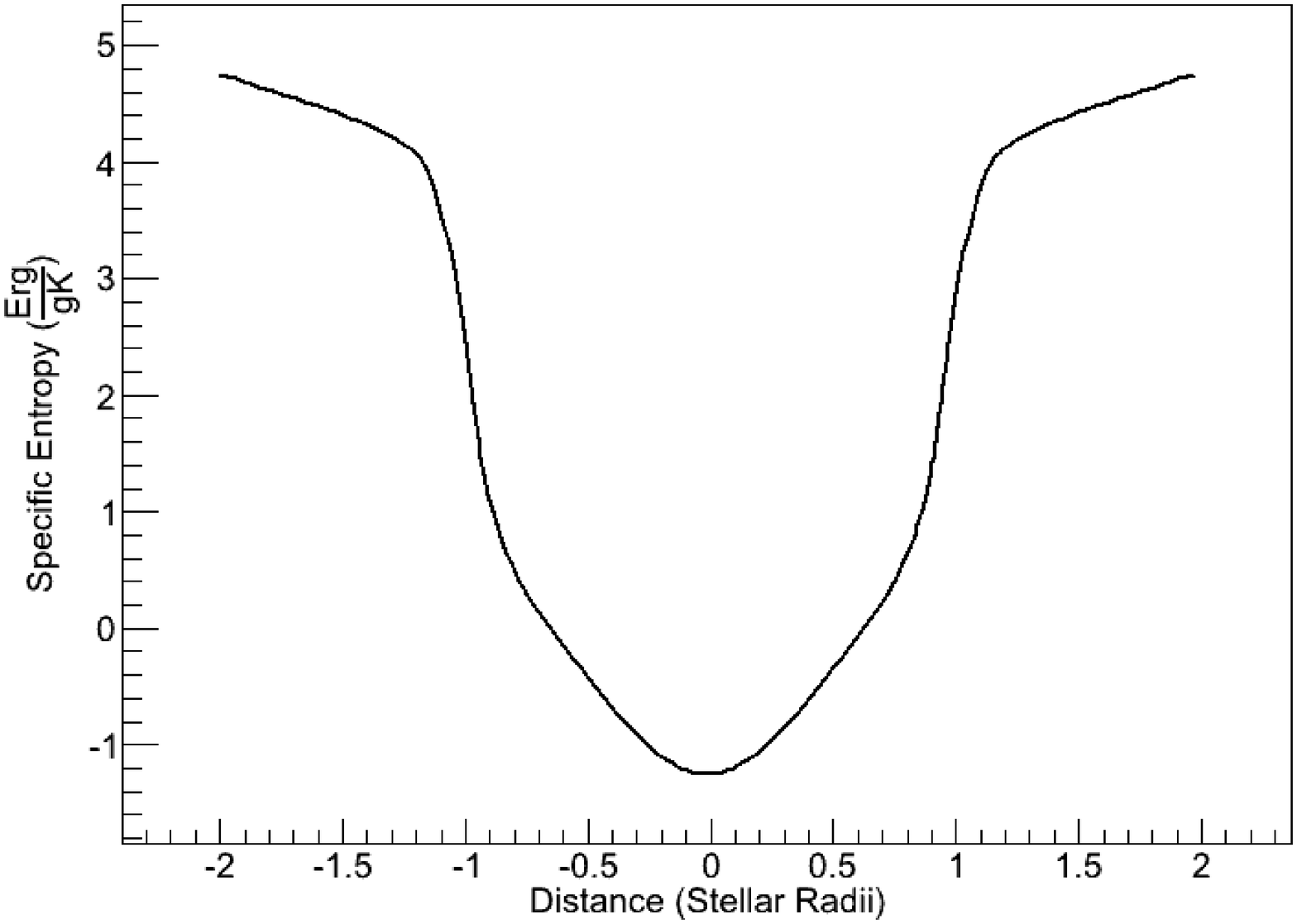}} 
\hfill 
 (b){\includegraphics[width=2.2in]{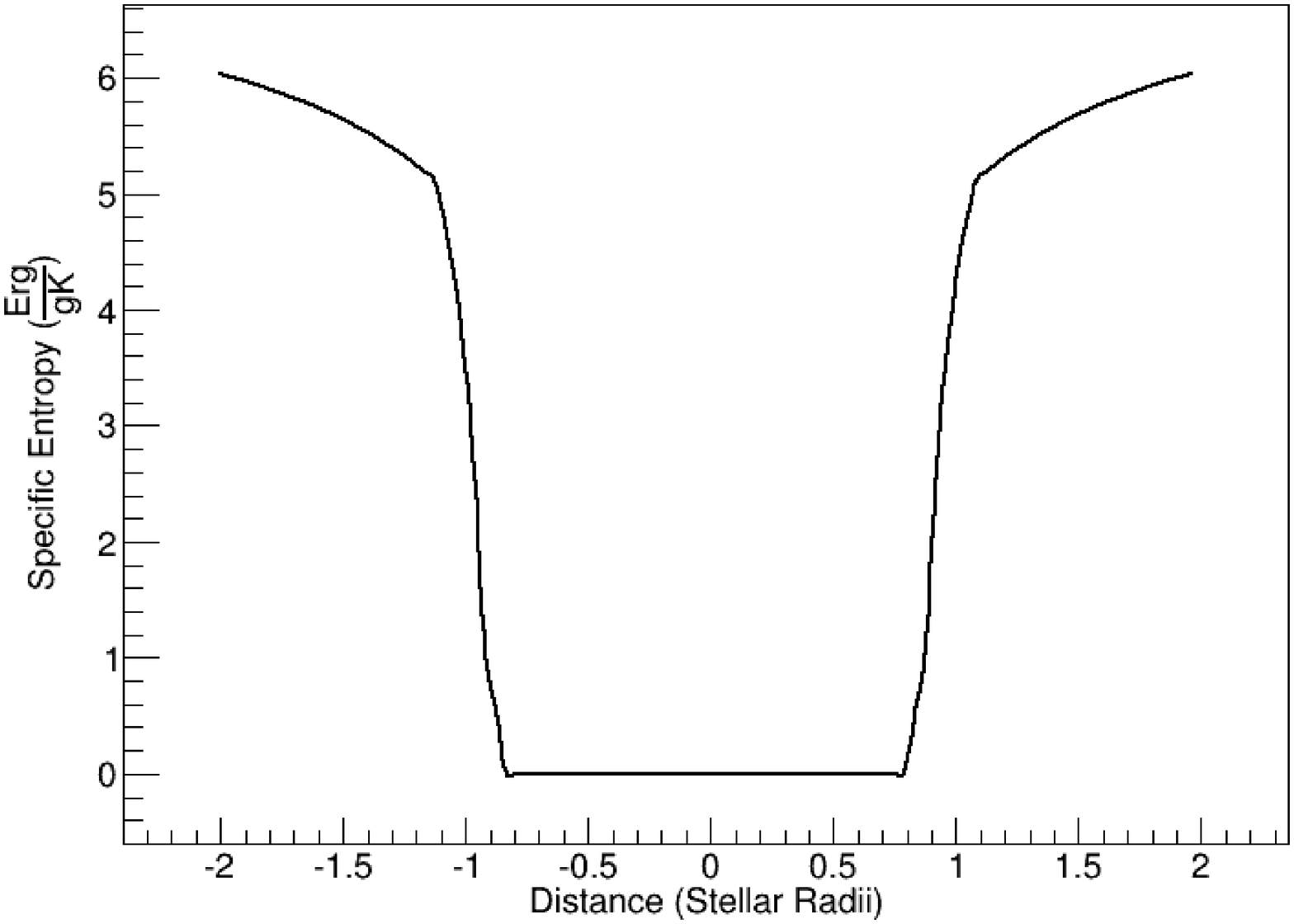}} 
 \caption{The specific entropy along the $x$-axis through the computational box for both models.  Panel (a) is for the polytrope of polytropic index $n=3$ the stably stratified star of non-constant specific entropy.  Panel (b) is for the polytrope of polytropic index $n=1.5$.  Note the star in this model is barotropic, as the specific entropy is uniform within the star.}
   \label{entss}
\end{figure}

To search for stable MHD equilibria in barotropic stars, we use the stagger code of \cite[Nordlund \& Galsgaard (1995)]{Nordlund_1995}, a three dimensional high order finite-difference MHD code in Cartesian coordinates.  The  model contains a perfectly conducting spherical star of radius one quarter the size of the grid box embedded in a poorly conducting atmosphere.  Two different stars were used in our simulations.  The first is an $n=3$ polytrope, which is stably stratified.  This model was used as a comparison to the second star, a polytrope of polytropic index $n=1.5$, which yields an initially uniform specific entropy, $s=0$ in the star, and thus a barotropic star.  The profiles of the specific entropies of both models can be seen in Fig. \ref{entss}.

Due to heat diffusion present in the code, the star will not keep a uniform specific entropy, resulting in an entropy gradient and a component of stable stratification.  To remedy this, we included a term in the energy equation in the code which forces the specific entropy back to its initial value.  The term takes the form :
\begin{equation}
\label{eq1}
\frac{de}{dt}=...+\frac{\rho T (s_0-s)}{\tau_{s}},
\end{equation}
where $\rho$, $T$, $s$, and $s_0$ are the density, temperature, specific entropy, and initial specific entropy respectively, and $\tau_{s}$ is the timescale at which this entropy term forces the star back to its initially barotropic structure.  In our models $\tau_{s}$ is a free parameter that was varied in different simulations with the only restriction being that it must be longer than the dynamical timescale of the star, and shorter than the diffusion timescale of the star.  With this term included, we are able to force the star to its initial barotropic structure and search for the existence of stable MHD equilibria.

\section{Results}
\begin{figure}[b]
\begin{center}
 \includegraphics[width=3.4in]{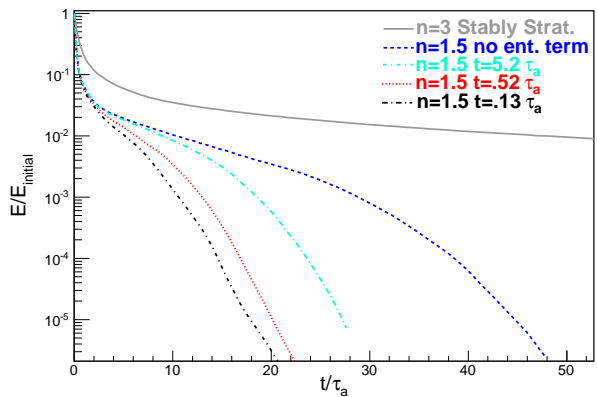} 
 \caption{Total magnetic energy relative to initial value versus time for differing models all starting from an initially random magnetic field configuration.  The solid gray curve is for a stably stratified model, which reaches a stable MHD equilibrium.  All other curves are of a barotropic star with differing values of the $\tau_{s}$, which are 0.13, 0.52, and 5.2 $\tau_{A}$ for the black dashed-dotted, the red dotted and teal dashed-dotted curves respectively as well as the blue dashed curve which does not include the entropy forcing term.  Note that none of the barotropic models reach a stable MHD equilibrium.}
   \label{fig1}
\end{center}
\end{figure}

All our models start out with an initially random magnetic field.  A series of simulations were run for the $n=1.5$ polytrope model with differing values for the entropy timescale, $\tau_{s}$, presented in Eq. \ref{eq1}.  We used four different $n=1.5$ models, one that did not include the entropy term in the energy equation, and three that included the entropy term with timescale values of 0.13, 0.52, and 5.2 $\tau_{A}$, where $\tau_{A}$ is the Alfv\'{e}n crossing time of the star defined as $\tau_{A}=R \sqrt{M/2E}$, where $R$ is the radius of the star, $M$ is the mass of the star and $E$ is the total magnetic energy in the star.  To compare to the stably stratified star, we also present one run in which an $n=3$ polytrope was used.  The evolution of the magnetic energy can be found in Fig. \ref{fig1}.  The main thing to take away here is that the stably stratified model reaches an equilibrium, as the field energy decays on a very long timescale, while all the barotropic models do not reach a stable equilibrium.  This suggests that stable stratification does seem to be important for reaching a stable MHD equilibrium.  This is further elucidated by looking at how quickly the magnetic energy decays for different values of the entropy timescale.  The smaller values of the entropy timescales keep the stellar structure very close to the initial barotropic structure throughout the evolution, but as this timescale is increased, or the entropy term in Eq. \ref{eq1} is not included for the $n=1.5$ stars, the decay of the magnetic energy is much slower.  This is due to the fact that the structure of these models obtain a slight gradient in their specific entropy profiles and thus a small amount of stratification, which act to slow the decay of the magnetic field slightly.

\section{Discussion}
We have searched for stable MHD equilibria in barotropic stars with an initially random magnetic configuration.  Our results indicate that from the random magnetic configuration used, no stable MHD equilibrium is reached.  This does not mean that stable MHD equilibria cannot exist in barotropic stars, but that starting from the particular random field configuration used, such an equilibrium cannot be reached.  Further study is needed, different initial magnetic field configurations should be studied, both from random magnetic field configurations, and from the physically motivated equilibria found in the axially-symmetric configurations of \cite[Armaza et. al. (2013)]{Armaza_2013}.

We note that the finite time scale $\tau_{s}$ can in fact represent an astrophysically relevant time scale. In entropy-stratified fluids such as the radiative regions of main-sequence stars and white dwarfs, a magnetic flux tube in mechanical equilibrium will have a lower specific entropy than the surrounding gas, but this entropy difference will be slowly erased by heat conduction. In the compositionally stratified neutron star cores, such a flux tube will have a higher fraction of charged particles (relative to neutrons) than its surroundings, but this composition difference can be erased by beta decays or ambipolar diffusion (\cite[Reisenegger (2009)]{Reisenegger_2009}; \cite[Hoyos, Reisenegger \& Valdivia (2008)]{Hoyos_2008}; and \cite[Reisenegger (2013)]{Reisenegger_2013}). Thus, all these stars behave as stably stratified (non-barotropic) on short time scales, but barotropic on longer time scales, which are controlled by (but substantially longer than) $\tau_{s}$.

\end{document}